\documentclass[pageno]{jpaper}

\usepackage[normalem]{ulem}

\usepackage{color}
\usepackage{soul}

\usepackage{blindtext}
\usepackage{epsfig}
\usepackage{syntonly}
\usepackage{rotating}
\usepackage{amsmath}
\usepackage{setspace}
\usepackage{verbatim} 
\usepackage{amssymb}
\usepackage{amsmath}
\usepackage{graphicx}
\usepackage{multirow}
\usepackage{colortbl}
\usepackage{enumerate}
\usepackage{wrapfig}
\usepackage{url}
\usepackage{algorithm2e}
\usepackage{cleveref}

\usepackage{ifthen}
\usepackage[dotinlabels]{titletoc}
\usepackage{colortbl}
\usepackage{array}
\usepackage{xcolor}
\usepackage{authblk}

\usepackage{tabularx} 

\definecolor{xxxcolor}{rgb}{0.8,0,0}
\definecolor{commentgrey}{rgb}{0.5,0.5,0.5}
\newcommand{\TODO}[1]{{\color{xxxcolor}{[}\emph{#1}{]}}}

\newcommand{\IGNORE}[1]{}

\newcommand{\AP}[1]{{\color{violet}{[}\textbf{AP: #1}{]}}}
\newcommand{\XY}[1]{{\color{blue}{[}\textbf{XY: #1}{]}}}
\newcommand{\SR}[1]{{\color{teal}{[}\textbf{SR: #1}{]}}}

\author[*]{Xuan Yang}
\author[*]{Jing Pu}
\author[*]{Blaine Burton Rister}
\author[*]{Nikhil Bhagdikar}
\author[*]{Stephen Richardson}
\author[$\dag$]{Shahar Kvatinsky}
\author[*]{Jonathan Ragan-Kelley}
\author[*]{Ardavan Pedram}
\author[*]{Mark Horowitz}
\affil[*]{Stanford University}
\affil[$\dag$]{Technion – Israel Institute of Technology}

\begin{document}

\title{A Systematic Approach to \\ Blocking Convolutional Neural Networks}
\date{}
\maketitle
\thispagestyle{empty}

\begin{abstract}

Convolutional Neural Networks (CNNs) are the state of the art solution for many computer vision problems, and many researchers have explored optimized implementations. Most implementations heur\-istically \emph{block} the computation to deal with the large data sizes and high data reuse of CNNs.
%
%
This paper explores how to block CNN computations for memory 
locality by creating an analytical model for CNN-like loop nests. 
Using this model we automatically
derive optimized blockings for common networks that improve the energy efficiency of custom hardware implementations by up to an order of magnitude. 
%
Compared to traditional CNN CPU implementations based on highly-tuned, hand-optimized BLAS libraries,
our x86 programs implementing the optimal blocking
reduce the number of memory accesses by up to 90\%.


\end{abstract}



\section{Introduction}
Convolutional Neural Networks~(CNNs)~\cite{krizhevsky2010convolutional,jarrett2009best,turaga2010convolutional,strigl5452452} 
have shown an ability to solve a number of challenging problems, 
giving rise
to a strong interest in their implementation, including customized hardware~\cite{
   liu2015pudiannao,
  chen2014dadiannao,
  jin2014teradeep,
  pham2012neuflow,
  %
  peemen2013memory}.
%
Customized hardware makes sense because CNN applications have high locality and high compute intensity, 
%
traits that custom hardware leverages.
%
Unfortunately, the working set of these applications is large: model parameters can be tens of megabytes, and the input data to a layer often consists of tens to hundreds of images. Thus, the design of the memory hierarchy and how the data is choreographed has a dramatic effect on the energy required for the computation.  

To achieve the desired locality, the problems must be partitioned into a number of smaller pieces, to allow these pieces to be stored in smaller memories close to the compute units.  The partitioning of the problem into the optimal set of smaller sub-blocks, or \emph{optimal blocking}, has been well studied for matrix multiplication~\cite{Gunnels:2001:FHM:645455.653765,Goto:2008:AHM:1356052.1356053} and is the problem we address in this paper for CPU and custom hardware implementations of CNNs.

Early attempts
\cite{
  Caffe,
  abuzaid2015caffe,
  DBLP:journals/corr/Lavin15,
  bergstra2010deep}
to optimize CPU and GPU CNN implementations treated the convolutional layers as matrix multiplication and used an optimized BLAS mat\-rix-mat\-rix-mul\-ti\-pli\-ca\-tion  (GEMM) routine~\cite{dongarra1990set}.
While GEMM implementations often perform
optimal blocking for matrix multiplication, we show that encoding convolution as matrix multiplication loses some of the locality of the original problem,  
results in replicated values, and significantly increases required memory accesses.

%
%
 GPUs don't really have a memory hierarchy to speak of; their combined register file is nearly as large as their last level cache. Thus blocking for a GPU is not about fitting data into the last level cache, but mainly about improving the concurrency, which is not the focus of this work, so we don't evaluate GPU results in this paper. 
However, near the end of Section~\ref{sec:cusRes} we discuss implications of our work for future GPU design.

Special purpose hardware solutions, both ASIC and FPGA~\cite{pham2012neuflow,farabet2011neuflow,peemen2013memory,zhang2015optimizing},
block the CNN algorithm
directly across multiple levels of compatible memory hierarchy and feed the data to a custom compute datapath. Given the large possible parameter space---the computation is a 4 level loop nest (x, y, input channel, kernel) around a 2-D convolution---blocking needs to consider all possible loop splits and loop orders for each split. As a result, previous designers tend to consider a limited part of the solution space. For example, given the limited on-chip memory resources on many FPGA systems, prior research in this area 
has built models to explore blocking for a single level of on-chip memory 
\cite{peemen2013memory,zhang2015optimizing}. 

This paper makes three main contributions, all designed to improve CNN blocking:
\begin{itemize}
\item We create an analytical model for, and optimizer of, 
memory energy and traffic within a multi-level memory hierarchy for CNN-like loop nests, designed to find the optimal blocking for any arbitrary memory hierarchy.

\item 
We use this optimizer to jointly find the memory hierarchy and blocking that yields the most energy efficient solution for a CNN problem, or a set of CNN applications. 

\item
 Our results show that 
 directly blocking CNNs yields better locality than the standard method of convolution using GEMM.
 %
\end{itemize}


\section{Overview of the CNN problem }
\label{sec:Overview}

Advances in training deep, multi-layer networks 
have led to a resurgence of their use in many problem 
domains~\cite{hinton2006fast}. In computer vision, convolutional neural networks (CNNs) have recently displaced classical image processing and machine learning methods for state-of-the-art performance 
on ma\-ny tasks, particularly in recognition.

Counter to the classical, freely-connected model
common\-ly associated with the neural network metaphor,
\emph{convolutional} neural networks are characterized by a highly restrict\-ed structure in which the network is organized into a pipeline or DAG of ``layers,'' and most layers are defined to perform a \emph{convolution} on their inputs. In vision problems, these layers can be thought of as producing and consuming images, with their neurons organized into a regular 3D grid of pixels (with image dimensions $x$ and $y$, and
$c$ for color channels). From this perspective, a CNN is more clearly thought of as a specialized class of image processing pipelines, rather than as a biological neural model. The operations in this pipeline---convolution, local response normalization, pooling, and fully connected 
layers---corre\-spond to the different ``layers" used in the network. 



\begin{itemize}
\item 
A \emph{convolutional layer} (Conv) corresponds to a filter bank. In the standard case of 3D input and output, a convolutional layer maps a 
$C \!\times\! X \!\times\! Y$ 
input to a 
$K \!\times\! X \!\times\! Y$ 
output using K shift-invariant 3D stencils, where each stencil is of the size 
$F_w \!\times\! F_h \!\times\! C$ 
(i.e., a set of $K$ 3-dimensional convolutions).
These $K$
$F_w \!\times\! F_h\!\times\! C$ 
stencil coefficients are the ``weights'' of the convolutional layer.
Here, $(X,Y)$ and $(F_w,F_h)$ are the image and kernel width and height dimensions and both image and kernels have the same depth dimension, which we define as $C,$ or the number of channels. Typically the dimensions of the 
kernels are much smaller than the image dimensions.

\item
A \emph{local response normalization}~(LRN) layer normalizes (scales) the value of each input by the sum of squared values in its neighborhood.

\item
A \emph{pooling layer} performs a windowed reduction using some aggregation function 
(most commonly, $\mathit{max}$), 
decimating the input. This maps a 
$C \!\times\! X \!\times\! Y$ 
input to a 
$C \!\times\! X' \!\times\! Y'$ 
output, using a 2D stencil window of some size over the input within which the aggregation function is applied to produce a single output. Pooling and LRN layers have no learned parameters (weights).

\item
Finally, a \emph{fully connected layer}~(FC) is what is most commonly thought of within the neural network me\-ta\-phor: an $M$ to $N$ mapping where all $M$ inputs drive all $N$ outputs, with unique weights for every input/output pair. This corresponds to an $M \!\times\! N$ matrix-vector multiplication, and with $M \!\times\! N$ unique weights has far more weight data relative to the size of the layer inputs and outputs ($O(\mathit{input}\times\mathit{output})$) than a convolutional layer.


\end{itemize}


Most of the computational work in real CNNs, and most of the intermediate data bandwidth, is in the convolutional layers. 
%
%
Meanwhile, the fully-connected layers~(Sec.~\ref{subsec:soa})
perform more work and load more parameters per input or output, but they are most commonly used at the end of a network pipeline, by which point the input has been heavily decimated.
The output of each layer may also be fed through a nonlinear \emph{activation function}. Since these are typically local point-wise arithmetic operations which can be easily computed, 
they only have a small influence on computation cost
and do not affect blocking (communication or locality) at all.

\subsection{CNN Characteristics}
\label{subsec:soa}

Current state-of-the-art networks for object recognition applications range from the order of ten 
layers to dozens.
The AlexNet architecture~\cite{NIPS2012_4824} has five convolutional layers with window sizes of 11$\times$11, 5$\times$5 and 3$\times$3 interleaved with several local response normalization layers, pooling layers, and followed by two fully-connected layers.
The VGGNet architecture~\cite{simonyan2014vggnet} comprises several different network substructures, each composed of many convolutional layers with 3$\times$3 filter windows, interleaved with pooling layers, and followed by two FC layers.
We focus our evaluation on these networks,
as well as the suite of applications demonstrated on recently published 
CNN hardware~\cite{chen2014diannao,chen2014dadiannao}. 
Table~\ref{tab:net_breakdown} shows the computation and memory breakdown for AlexNet and two types of VGGNet architecture. From the table we can see 
that
Conv layers are the most computationally intensive layer 
while FC layers consume the most memory.

\begin{table}[ht]
  \begin{centering} 
    \begin{tabular}{|l|c | c|}
      \hline 
      \bf
      & \bf  MACs$\times 10^{9}$ 
      & \bf  Mem (MB) \tabularnewline
      \hline 


AlexNet Convs   & \hphantom{0}1.9	& \hphantom{0}2 \tabularnewline
VGGNet-B Convs  & 11.2              & \hphantom{}19 \tabularnewline
VGGNet-D Convs	& 15.3	            & \hphantom{}29 \tabularnewline
\hline
AlexNet FCs	    & 0.065	& 130 \tabularnewline
VGGNet-B FCs	& 0.124 & 247 \tabularnewline
VGGNet-D FCs	& 0.124	& 247 \tabularnewline
      \hline 
    \end{tabular}
    \par
  \end{centering}
  
  \protect\caption{Computation (measured in number of multiply and accumulate operations) and memory consumption breakdown of state-of-the-art networks (each pixel and coefficient is 16 bits).
  }
  \label{tab:net_breakdown}
\vspace{-8pt}
\end{table}

\subsection{Related Work}
\label{sec:Related_work}

To achieve extremely high energy efficiency, a number of recent efforts have proposed specialized architectures for CNN workloads. The DianNao family of architectures was built around a customized inner-product unit designed for 
CNNs
and other machine learning 
algo\-ri\-thms.
In its first instantiation,
computation was minimally tiled to fit into a single level of small buffers~\cite{chen2014diannao}.
In a later iteration, the original unit was surrounded by eDRAM large enough to store the complete coefficient and data sets (assuming all network coefficients can fit),
but no further blocking was performed~\cite{chen2014dadiannao}.

The NeuFlow architecture builds CNNs in a systolic array such that each processing element communicates only with its neighbors, with results streaming to and from DRAM~\cite{pham2012neuflow,farabet2011neuflow}.
NeuFlow designs have only been implemented as FPGAs, not ASICs, where the flexibility to customize the memory hierarchy for energy efficiency is limited.
Its successor, the TeraDeep architecture, uses a fixed blocking strategy for convolutional layers~\cite{gokhale2014teradeep,dundar12teradeep,jin2014teradeep}.
Both Peemen et al.~\cite{peemen2013memory} and 
Zhang et al.~\cite{zhang2015optimizing} 
explored blocking in an FPGA CNN accelerator,
but considered only two levels of memory and blocking, and sought only to minimize off-chip bandwidth, not total memory energy. Section~\ref{sec:cusRes} shows how using a better blocking improves the energy of
these custom systems. 




Most CPU and GPU implementations for CNN have used some combination of hand-tuned GPU kernels~\cite{cuDNN,NIPS2012_4824}, 
optimized Basic Linear Algebra Subprogram routines (BLAS) \cite{Caffe}, and metaprogramming~\cite{bergstra2010deep}. These implementations generally rely on coarse-grained blocking across images to expose opportunities for locality, and to improve arithme\-tic intensity, which is exploited by using optimized GEMM kernels and similar code.
However, in order to utilize the optimized GEMM kernels, these implementations have to first remap 3D input tensors to 2D matrices, a process called \textit{lowering}.


 An early attempt called {\it Caffe}~\cite{Caffe}
 used these implementations and achieved good results, but was found to be suboptimal, since lowering
 duplicates much of the data and wastes memory,
 and, specifically in Caffe, lowering happens distant from the computational core. 
 Caffe con Troll~\cite{abuzaid2015caffe} is an improvement that tiles the original image before applying the lowering process, there\-by reducing the memory resource waste. It also does better batching, with each core working independently on a different group of images.

While at first sight the high level looping for convolution and GEMM seem similar, there are some fundamental differences. 
First and foremost, the nature of low-level computations in convolution and GEMM are different because of the extra axes of reduction, and because of the overlaps in the sliding convolution windows that provide another level of temporal locality GEMM does not have.
Second,
the blocking of convolution affects the amount of refetches to overlapping regions of blocked tiles, which does not exist in GEMM blocking.
Third, the 3D composition of the data stream
implies a larger space of possible blocking than 2D matrix, 
and therefore the search for optimal blocking is more complex.
By taking advantage of the temporal locality the overlapped windows provided, and reducing recomputes or refetches at the boundary region, directly blocking convolution should achieve better performance and memory usage. These differences are explained in more detail in the next section.

Based on the nature of Conv and FC layers,
we propose a general way of analyzing and optimizing the blocking of a CNN onto a memory hierarchy, which can be applied to any 
convolutional neural network on existing and custom architectures with a memory hierarchy.
We further generalize this to co-optimize the blocking with the design of the memory hierarchy, creating
more energy-efficient solutions.

Existing approaches to this problem include {\it polyhedral algorithms}~\cite{pouchet2013polyhedral, bondhugula2008practical, bondhugula2014tiling, zhang2015optimizing, bondhugula2016pluto+} and {\it cache oblivious algorithms}~\cite{frigo2005cache, strzodka2010cache}.  Polyhedral algorithms are powerful for optimizing both well-aligned and irregular complex loop nests, but because of its more complicated data structure, it usually requires significant computation. 
Cache oblivious algorithms are designed to use cache optimally in an asymptotic sense, and the optimization is independent of specific machines or cache sizes. Our framework leverages the advantages of both 
approaches.  Like Polyhedral algorithms, we can optimize to specific memory hierarchies, avoiding poor fitting to specific cache sizes that cache oblivious algorithms have. Like cache oblivious algorithms our evaluation procedure is low complexity, allowing us to consider a large number of memory configurations rapidly.


\begin{figure*}[t] 
   \centering
   \includegraphics[width=.95\textwidth]{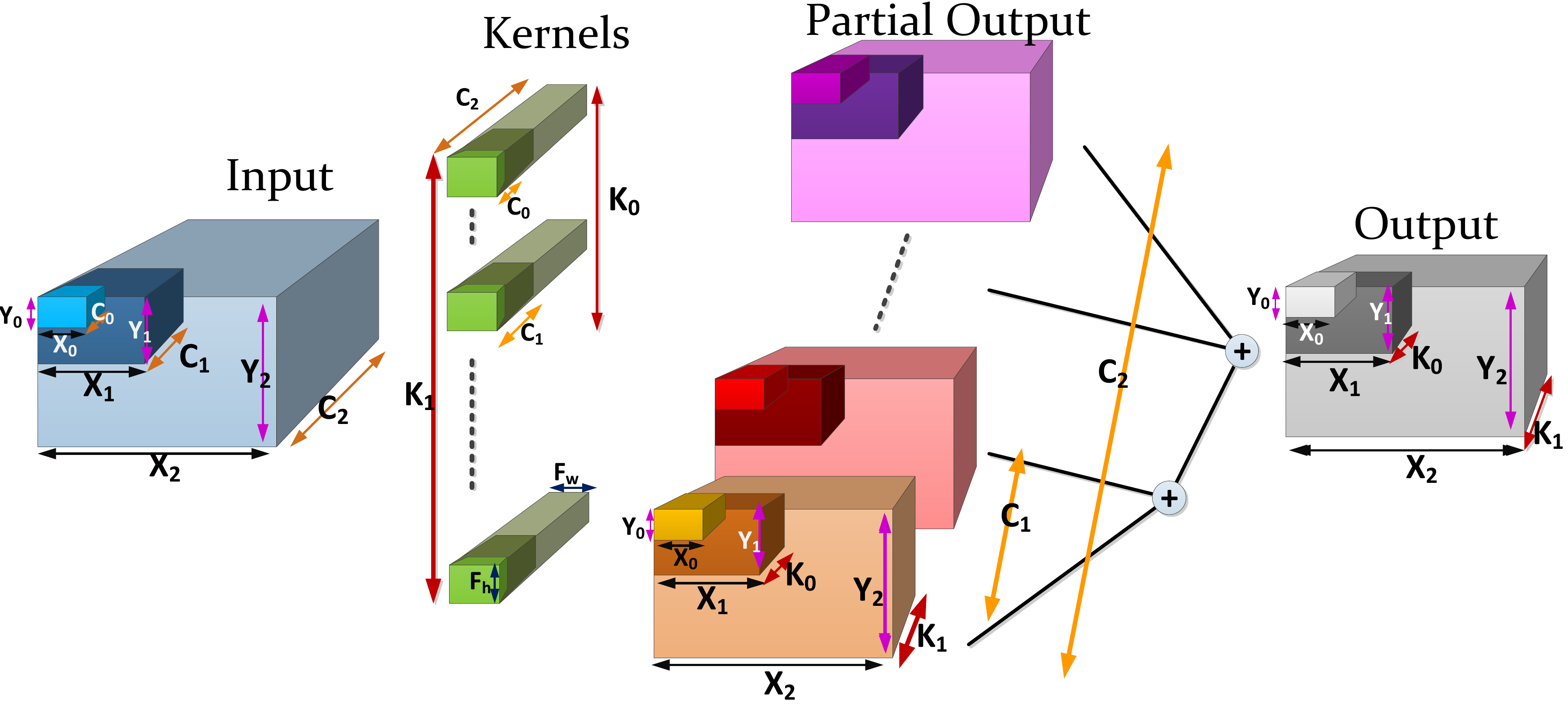}
   \caption{Hierarchical blocking of a single convolutional layer. The six-dimensional overall problem domain ($X,Y,C,F_w,F_h,K$) depicted in Figure~\ref{Alg:sim_alg} is blocked to three levels in the input domain ($\{X,Y,C\}_{\{0,1,2\}}$), and two levels in the set of kernels ($K$) which correspond to the third dimension of the output domain ($\{X,Y\}_{\{0,1,2\}}, \{K\}_{\{0,1\}}$). Partial results for each output pixel are accumulated hierarchically across the three levels of blocking in $C$.  
   }  
   \label{fig:hierarchy}
\end{figure*}

\begin{algorithm}[!t]
\vspace{1\baselineskip}
\LinesNumbered
\footnotesize
\SetAlCapNameSty{textbf}
\SetAlgoCaptionLayout{small}

\vspace{.5\baselineskip}
\nl \For{$k_0 = 0 : K$} {

 \vspace{.5\baselineskip}
 \nl \For{$c_0 = 0 : C$} {

  \vspace{.5\baselineskip}
  \nl \For{$y_0 = 0 : Y$} {

   \vspace{.5\baselineskip}
   \nl \For{$x_0 = 0 : X$}{

    \vspace{.5\baselineskip}
    \nl \For{$fh = -\frac{F_h-1}{2}:\frac{F_h-1}{2}$} {

     \vspace{.5\baselineskip}
     \nl \For{$fw = -\frac{F_w-1}{2}:\frac{F_w-1}{2}$}{
%
%
%
%
       \vspace{.30\baselineskip}
       $output[x_0, y_0, k_0] +=$
       \\ \vspace{.35\baselineskip}
       $input[x_0+fw, y_0+fh, c_0]\times$
       \\ \vspace{.35\baselineskip}
       $kernel[fw, fh, c_0, k_0]$ \; 
     }
    }
   }
  }
 }
}
\vspace{0.5\baselineskip}
\caption{The 6 nested loops of the 
convolutional layer. In this simple example, the loop order is represented by $F_wF_hXYCK$
}
\label{Alg:sim_alg}

\end{algorithm}

\section{Convolutional layer analysis}
\label{sec:Theory}

Each CNN layer processes two grids---the 3-D input image ($X, Y, C$) and the 4-D kernel weights 
$(F_w,F_h, C,$ $K)$~---to 
produce a single 3-D output grid 
($X, Y, K$). 
This computation is depicted in 
Figure~\ref{fig:hierarchy}. 
Convolution layers have high compute intensity, but, like GEMM, the large data sizes require the computation to be properly blocked for efficient execution.  What makes the blocking hard is the fact that all the data fetched---input, kernel, and output---is reused multiple times.  In a convolutional layer, output pixels at different input image positions share the same kernels; different kernels share the same input image; and overlapping windows share input pixels.  So any blocking scheme will cause some data to be refetched. 

Since the energy cost of a fetch depends on memory size, a good blocking minimizes memory energy by serving most of the data from small 
memories and
minimizing the amount of data that these memories need to fetch from larger, higher energy memories in the memory hierarchy. 

\subsection{Basic Blocking Notation}
The computation being performed by a convolutional lay\-er can be easily expressed as a 6 layer loop nest\footnote{Actually one can consider this problem to be a 7 level loop nest, since this computation is repeated over a number of images, and you sometimes want to block over images as well, especially for the FC layers.} as shown in 
Algorithm~\ref{Alg:sim_alg}.
Since there are no dependencies in this computation, the loops in the algorithm can be done in any order.  
We will represent a particular implementation order by creating a string that indicates the loop order from  innermost to outer.  Thus $F_wF_hXYCK$ presents the computation shown in Algorithm~\ref{Alg:sim_alg}. The value of each of the variables in our string represents the number of iterations done at this loop level.

Given this initial loop nest, blocking can be thought of as simply splitting a number of loops, and then exchanging the order in which these split loops are executed. In our notation, when the $X$ loop is split, $X_0$ represents the inner part of the $X$ loop and the value of $X_0$ represents the range of the data computed in this loop.  
For $X_0$, the loop variable $x_0$ increments by one, and the value of $X_0$ remains equal to the number of iterations in the inner loop. $X_1$ represents the outer loop and its value again represents the range of data computed in this loop.  
In this case, the loop variable $x_1$ increments by $X_0$, so the number of iterations is $X_1/X_0$.
Multi-level blocking occurs when a single loop is split multiple times, and is easy represented in our notation extending $X_1$ to $X_n$; the loop variable for $X_n$, $x_n$, increments by $X_{n-1}$ on each iteration.

Using this representation of nested loops, the blocking problem is easy to state:  
Find the loop order string,
and the size of each loop, which minimizes the memory energy.  This formulation also makes it easy to see the large space of possible blocking -- the number of possible loop orders is quite large, and for each loop order, we need to compute the loop sizes that minimize the memory energy.

\subsection{Memory Hierarchy}
To solve this optimization problem, we need to compute the memory energy for a given blocking ``string'' which will depend on the memory hierarchy present in the design.  While in the final design the input, kernel, and output data at each level of the memory hierarchy may be stored together, for this analysis it is convenient to think of them as separate memory structures. Thus we will consider a memory for kernel coefficients $\textit{KB}$ (kernel buffer), input image data $\textit{IB}$, and output data $\textit{OB}$.  Since these memories exist at multiple levels in the memory hierarchy, we use $\textit{KB}_0$, $\textit{IB}_0$, $\textit{OB}_0$, to indicate the kernel, input, and output memory that is closest to the compute unit, and each buffer at level $i$, (e.g. $\textit{IB}_i$) fetches its data from the buffer at level $i+1$ ($\textit{IB}_{i+1}$). 

To minimize the memory energy, we would like to fetch data from the smallest possible memory. This leads to a simple rule about where in the computation (at which loop nest) we should add a buffer: a buffer must be added anytime the added loop reuses the same data in its loop iteration. Thus adding buffers can be thought of as a recursive process. Assuming that we have already added buffers optimally up to level $i-1$:

\vspace{0.5\baselineskip}
\noindent
1) When a new X or Y loop $X_i,Y_i$ is added to the inner loops, a series of image blocks are streaming through the same set of kernels producing output images with dimension of $X_i,Y_i$. Therefore one can save those kernels that are being used $X_iY_i/(X_{i-1}Y_{i-1})$ times in a new kernel buffer.\footnote{Note that the outputs produced from each input are distinct, so the partial outputs are not reused for a given set of kernels.  They will be buffered at the level where their reuse occurs.} For maximum reuse of kernel coefficients with $X_i, Y_i$ image window size, the kernel buffer contains all elements that are used by the inner loops:\footnote{Note that this buffer size does not depend on the ordering of the inner loops.}  $\textit{KB}_i[F_w,F_h K_{i-1},C_{i-1}]$, assuming loops $F_w$, $F_h$ have been added before the current level. 

\vspace{0.5\baselineskip}
\noindent
2) When a new C loop $C_i$ is added, a series of images and kernels are streamed and $C_i$ channels reductions are being performed on the same set of outputs. Therefore those partial outputs are being reduced $C_i/C_{i-1}$ times, and should be stored in a new output buffer to prevent these fetches from going to a larger memory at a higher level in the memory hierarchy.
For maximum reuse of kernel partial outputs,  the output buffer contains all elements that are computed by the inner loops: $\textit{OB}_i[ X_{i-1},Y_{i-1},K_{i-1}$].

\vspace{0.5\baselineskip}
\noindent
3) Finally, when a new K loop $K_i$ is added to the existing inner loops, each iteration will load a new kernel, but each of these kernels will operate on the same set of input data producing $K_i$ channels of the output. 
Therefore one can save those input data that are being used $K_i/K_{i-1}$ times in a new input buffer. 
To achieve the maximum reuse of data with $K_i$ kernels, the input buffer must contain all the input elements that are used by the inner loops:  $\textit{IB}_i[ X_{i-1},Y_{i-1},C_{i-1}]$. 
\vspace{0.5\baselineskip}

For level 0, the loop variables are the same as
in Algorithm~\ref{Alg:sim_alg}, but with loop sizes being $F_w$, $F_h$, $X_0$, $Y_0$, $C_0$, $K_0$ from inner to outer loop. The way to compute the buffer sizes and buffer reuse still apply, with $X_{-1}$,~$Y_{-1}$,~$C_{-1}$,~$K_{-1}$ $=1$. 

Note that loops $F_w$, $F_h$ are not required to be the innermost loops. With buffers optimally up to level $i-1$, when a new $F_w$ or $F_h$ is added to the existing inner loops, each iteration will load a new coefficient from the next position inside the window.  Those coefficients will operate on the same set of input data and reduce to the same set of output data.  Therefore, one can save those input images and output images, as input images are being used $F_w$ or $F_h$ times and output images are being used $2F_w$ or $2F_h$ times. The set of input and output data should be stored in a new input and output buffer respectively to prevent these fetches from going to a larger memory at a higher level in the memory hierarchy.

Figure~\ref{fig:hierarchy} demonstrates two levels of nested blocking for each dimension, and the associated buffers. The inner loop takes a small amount of input data with block size $X_0Y_0C_0$ and convolves it with $K_0$ kernels to create some partial outputs with block size $X_0Y_0K_0$. A complete output cannot be generated until all the channels of the input are processed for that kernel and the output pixel is generated, which will happen only when all of the channels ($C_2$ loop) finish.

\subsection{Coarse-grain Parallelism}
\label{sec:Parallel}

\begin{figure}[!tb] 
   \centering
   \includegraphics[width=0.45\textwidth]{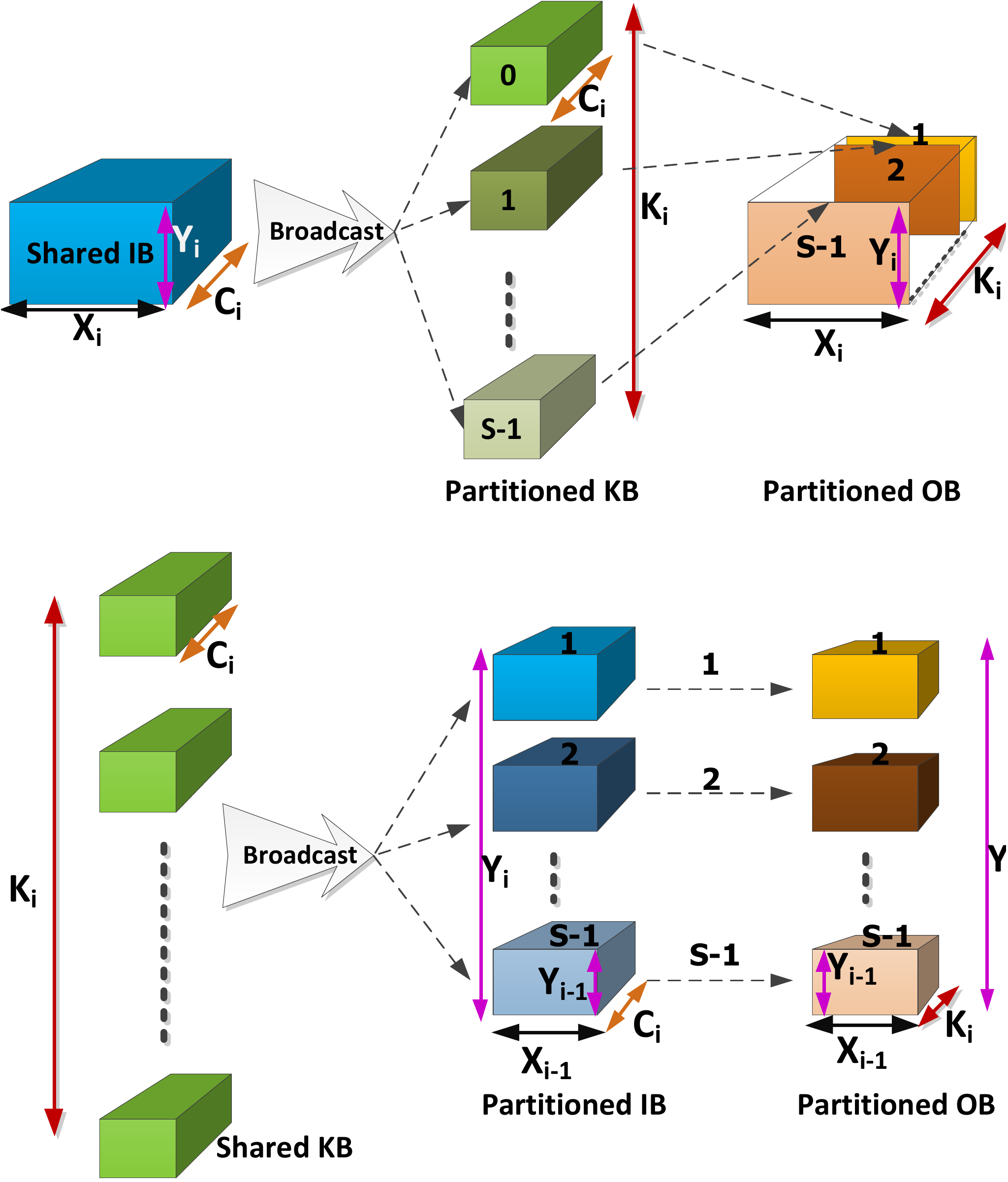} 
   \vspace{5pt}
   \caption{Multicore partitioning. Top: kernel partitioning broadcasts a shared input to separate cores, each of which processes a disjoint subset of the kernels to produce a disjoint slab of the output (in the $K$ dimension). Bottom: input partitioning broadcasts all kernels across cores which each process a different subset of the input to produce a disjoint subset of the output, shown here in the $Y$ dimension.}
   \label{fig:multi-core}
   \vspace{5pt}
\end{figure}

Multi-core parallelism can improve throughput and area efficiency by sharing the large, higher levels of the memory hierarchy. 
Fortunately, this type of parallelism is easily integrated into our blocking framework as a physical unrolling of an outer loop in the blocking string. This section describes the ramifications of this unrolling which, like memory hierarchy, depends on which loop (X, Y, C, K) is unrolled.

Suppose we apply parallelism for $S$ cores at a given level $p$ by unrolling 
that loop $p$ 
across the processors. The first constraint is that we need to block the application such that the dimension 
being unrolled, e.g. $C_p$, is $S$ times that of the previous level, $C_{p-1}$.
The parallelism can be performed by partitioning the problem across the input XY, the kernels K, or the channels C.  It turns out that partitioning across C 
requires extra communication between the processors to reduce the partial products of output~(see Figure~\ref{fig:hierarchy}) and hence it is not considered further. 

Like the analysis for the memory hierarchy, the critical issue to consider is which memory buffer is being refetched in this loop, since, by unrolling, this sharing becomes effectively a parallel broadcast.  The parallel broadcast obviates  the need to add a buffer at this level, but this fetched data must be sent to all the computing units.  The other buffers can be partitioned onto the different processor units. We consider two options:

{\bf K partitioning,} unrolling a K outer loop: This unrolling effectively
partitions the KBs and distributes them among the $S$ cores. Now the ``refetched'' input is broadcast to all the cores in the system, and 
each core is applying the same image to its own dedicated set of kernels and is producing its own dedicated output set of channels.  Thus in this architecture the current $KB$ and $OB$ should be partitioned into each core, and $IB$ should remain global to send data to all the cores. This is shown in the top of Figure~\ref{fig:multi-core}.



{\bf XY partitioning,} unrolling an X or Y loop:
Since the kernel data is being reused, IB and OB are distributed among $S$ cores (partitioned across X or Y) and KB is globally available to all cores. This is shown in the bottom of Figure~\ref{fig:multi-core}.

A multi-layer CNN problem has additional communication costs that need to be considered.  If the K loop is unrolled, at the end of a layer, different processors will hold output images for different kernels.  Since these outputs are the input channels for the next stage, we will need to eventually broadcast all of this data to all the processors.  For XY unrolling, if the next level uses the same XY unrolling, then this output data can remain ``local'' across the levels.
Therefore, XY partitioning is the most symmetric solution, where cores can successfully process data without large communication between layers as the computations are mostly independent.

\subsection{Memory Energy}
\label{sec:memEng}

We use a simple model to estimate memory energy: we sum the cost of all the memory fetches needed to complete the application.  Since the fetch order is set by the blocking, we can easily compute the number of references to each level in the memory hierarchy.  Once the number of references are known, we use estimates of the memory access energy as a function of memory size
using
%
Cacti~\cite{cacti_micro07}, as described in Section~\ref{sec:cusSol} and as summarized in Table~\ref{tab:table_energy}.

\begin{table*}
\begin{centering}
\vspace{-10pt}
\begin{tabular}{|c|c|c|c|}
\hline 
New Loop &Buffer Name & Buffer Size & Buffer Refetch Rate\tabularnewline
\hline 

$K_i$& $IB_{i}$ & $(Y_{i-1}+F_h-1)(X_{i-1}+F_w-1)C_{i-1}$ & $ (K_{i}(Y_{i-1}+F_h-1)(X_{i-1}+F_w-1))/(K_{i-1}Y_{i-1}X_{i-1})$\tabularnewline


$C_i$& $OB_{i}$ & $Y_{i-1}X_{i-1}K_{i-1}$ & $2C_i/C_{i-1}$\tabularnewline


$X_i$ or $Y_i$  &$KB_{i}$ & $C_{i-1}K_{i-1}F_hF_w$ & $(X_iY_i)/(X_{i-1}Y_{i-1})$\tabularnewline

\hline 
\end{tabular}
\par\end{centering}
\vspace{5pt}
\protect\caption{Buffer sizes and refetch rates for kernel, output, and input buffers at level $i$ of the blocking hierarchy, assuming that we have already added buffers optimally up to level $i-1$. Both the required buffer size and the refetch rate are determined from the domain required by all enclosing loops below the level at which the given buffer is allocated.}
\label{tab:equations}

\end{table*}

\begin{table}[th]
\centering 
\begin{tabular}{|r|r|r|r|r|r|}
\hline 
Size(KB) & 64\,bits & 128\,bits & 256\,bits & 512\,bits  \tabularnewline
\hline 
\hline 
1 & ~1.20 & ~0.93 & ~0.69  & ~0.57  \tabularnewline
\hline 
2 & ~1.54 & ~1.37 & ~0.91 & ~0.68 \tabularnewline
\hline 
4 & ~2.11 & ~1.68 & ~1.34 & ~0.90 \tabularnewline
\hline 
8 & ~3.19 & ~2.71 & ~2.21 & ~1.33 \tabularnewline
\hline 
16 & ~4.36 & ~3.57 & ~2.66 & ~2.19 \tabularnewline
\hline 
32 & ~5.82 & ~4.80 & ~3.52 & ~2.64  \tabularnewline
\hline 
64 & ~8.10 & ~7.51 & ~5.79 & ~4.67  \tabularnewline
\hline 
128 & 11.66 & 11.50 & ~8.46 & ~6.15 \tabularnewline
\hline 
256 & 15.60 & 15.51 & 13.09 & ~8.99  \tabularnewline
\hline 
512 & 23.37 & 23.24 & 17.93 & 15.76 \tabularnewline
\hline 
1024 & 36.32 & 32.81 & 28.88 & 25.22  \tabularnewline
\hline 
 $>$16384   & \multicolumn{4}{c|}{320}   \tabularnewline
\hline 
\end{tabular}
\vspace{5pt}
\caption{Memory access energy per 16 bits (pJ/16b) for various memory sizes and word lengths. For memory size in the range of 0.25KB to 16MB, we use SRAM. When the memory size exceeds 16MB, we use DRAM.}
\label{tab:table_energy} 
\end{table}

We compute the number of accesses to each memory by introducing 
refetch rate $\textit{RR}_i,$ 
the number of times a piece of data is fetched from a certain buffer after initially being loaded into that buffer (Table~\ref{tab:equations}). The number of fetches of data at 
memory hierarchy level $i$
is $\textit{RR}_i$ times the fetches at level $i+1$. With this definition, the total accesses of a buffer at memory level $i$ after running through the entire problem is the product of the refetch rates at higher levels in the hierarchy, which indicates the total number of times each data is fetched, multiplied by the number of elements in the top level memory, $\alpha$:

\vspace{-2\baselineskip}

\begin{small}
\begin{multline}
\hfill
\text{~~~~ total access of buffer at level $i$ ~ = ~ }
\alpha\times
\Pi_{j=i}^{n} RR_j 
\label{eq:accessbufferi}
\end{multline}
\end{small}

\vspace{-\baselineskip}

The final issue we need to address is to account for the energy cost of a memory fetch which broadcasts to multiple processors in a solution that uses coarse gain parallelism. In order to estimate this cost in a manner that could scale with technology, we use an indirect method to estimate the energy of the broadcast bus: 
%
we find a memory block of a size such that it has comparable energy to this broadcast.

How do we find the right size for this memory block?
We leverage the fact that large SRAMs are built from smaller memory arrays, so the energy increase as the memory gets larger is mostly from the energy to communicate the data 
to the output port from the array where it was stored. 
This communication cost is similar to the broadcast cost we are trying to estimate.
To find the size of this equivalent memory, we assume that the area of the cores will be dominated by the area of the last level memory. Thus the area that the data needs to be broadcast is the same as 
the size of the total embedded memory of the design.
As a result, we can estimate the broadcast cost by the fetch energy of a single memory of this size.  

While this method can
be tuned by allocating some area for the processor, since the energy of a memory reference is a
weak function of the cache size, this simple approximation seems to work well for the examples we have explored.

\subsection{Optimization Framework} \label{sec:optimization}


With our analysis of the optimal memory hierarchy and our memory model, we can compute the memory energy for any given blocking string. From the string we first compute the size and number of levels of memory in the design, determine the number of accesses each memory will need to serve, and then estimate the total energy required.  Unfortunately this problem is not convex, so finding a true optimum requires exhaustive search. With six dimensions and multiple levels to block for each layer, the design space becomes large but not computationally intractable. 
Knowledge about
blocking CNNs allows us to trim this space, 
further reducing
its size. For 2-level blocking there are approximately 3000 strings that need
their parameters optimized;
for four levels, the number of strings are in the order of a million. While four levels of memory hierarchy might seem excessive, it is important to remember this blocking scheme can block for registers as well as memories, and includes the DRAM as the final level in the hierarchy, which increases the number of levels in the memory hierarchy.

Our initial optimizer simply enumerated all consistent parameter values in all possible strings and chose those with minimum energy. While this was computationally expensive, it was still feasible to optimize a single 4-level 
layer (the optimization took around 24 hours on a Xeon E5645 processor using a single thread).

To optimize
a CNN layer for a fixed memory hierarchy, for each string we continue to pack the lower level buffers into the lowest available level of memory hierarchy, always adding the unpacked buffer with the highest number of accesses.  When the current memory level does not have enough remaining space to fit the added buffer, we place that and all subsequent buffers into the next level of the memory hierarchy until it becomes full.

We use two characteristics to speed up the optimization procedure 
enough to allow
work on multi-layer optimization. First, we notice that the computation time grows exponentially with the number of memory levels (string length), so short strings are much easier to optimize than long strings. Second, we notice that while the blocking of level $i$ strongly depends on the blocking of level $i+1$, the effects of blocking levels $i+2$ and $i+3$ are much smaller. The blocking at level $i$ is trying to minimize the energy cost of doing all the fetches it needs from the local memory plus the memory fetches needed to load this memory, so it needs to know the energy costs of the load accesses (to level $i+1$). If this memory has reasonable refetch rates, which it should, then the component of the load cost that depends on the higher level decisions should be small.

We speed up the optimization by iteratively optimizing the blocking from lower memory levels to higher ones, corresponding to optimizing from inner to 
outer loops. Conceptually we start by optimizing a two level hierarchy, and then in each iteration, we add blocking into the higher memory hierarchy, and reoptimize the lower blocking levels. 
Since adding a new level of blocking may affect the inner loops and cause them to become suboptimal, we introduce some randomness to the inner loops before beginning each new iteration. First, rather than choosing the lowest energy design, the best 128 loops are used as seeds for the next level. Next, additional seed strings are created for the optimized inner loops by randomly perturbing the loop sizes and exchanging some adjacent loops. Each of these seeds are then used to search for the optimal $i+1$ level blocking. The resulting procedure can complete a 4-5 level optimization in a few minutes, and the energy per operation for the first five benchmarks
in Table~\ref{tab:table1} 
is higher than those found by full enumeration, but only by 8\% or less.


\subsection{
Flexible memory design 
}


In most applications, multiple layers, or even multiple problems, often need to run on the same system. Thus, while the optimization procedure described in the previous section is informative, it does not directly solve the problem of optimizing the blocking of a complete multi-layer CNN.  


With the objective of minimizing 
energy per op of each layer and total energy consumption of all the layers, we perform the optimization in two steps. First, we explore the energy and area design space for each layer separately, leveraging the single layer optimizer working on a given memory hierarchy. After the first step, each layer will record a set that contains its 10 most energy efficient design points still under the area budget. The second step is to find common design points among those sets to optimize total energy consumption.

This optimization runs with a fixed number of memory levels.  If one wants to optimize the number of memory levels, multiple runs are required. 
\section{Methodology }
\label{sec:method}


\begin{table}
\centering 
\begin{tabular}{|l|r|r|r|r|r|r|r|}
\hline 
Layer & X & Y & C & K & $F_w$ & $F_h$ \tabularnewline
\hline 
\hline 
Conv1~\cite{NIPS2012_4824}& 256 & 256 & 256 & 384 & 11 & 11 \tabularnewline
\hline 
Conv2~\cite{farabet2011neuflow}& 500 & 375 & 32 & 48 & 9 & 9 \tabularnewline
\hline 
Conv3~\cite{sermanet2011traffic}& 32 & 32 & 108 & 200 & 4 & 4 \tabularnewline
\hline 
Conv4~\cite{simonyan2014vggnet}& 56 & 56 & 128 & 256 & 3 & 3\tabularnewline
\hline 
Conv5~\cite{simonyan2014vggnet}& 28 & 28 & 256 & 512 & 3 & 3 \tabularnewline
\hline 
\hline 
FC1~\cite{sermanet2011traffic}& - & - & 200 & 100 & - & -  \tabularnewline
\hline 
FC2~\cite{simonyan2014vggnet}& - & - & 4096 & 4096 & - & -  \tabularnewline
\hline 
Pool~\cite{simonyan2014vggnet}& 56 & 56 & 128 & - & 2 & 2 \tabularnewline
\hline 
LRN~\cite{NIPS2012_4824}& 55 & 55 & 96 & - & - & -  \tabularnewline
\hline 

\end{tabular}
\vspace{5pt}
\caption{Problem dimensions of the benchmark network layers we used in our evaluation.}
\label{tab:table1} 
\end{table}

To evaluate the effectiveness of our approach to 
blocking,
we apply it to both software CNN implementations running on processors with conventional memory hierarchies, and custom hardware implementations where we can create an optimized memory hierarchy. 
Table~\ref{tab:table1} lists the benchmarks 
we use.
The majority of benchmarks are chosen from state-of-the-art convolutional neural networks such as AlexNet~\cite{NIPS2012_4824} and VGGNet~\cite{simonyan2014vggnet}. Others are chosen from networks that have been mapped to custom hardware by other researchers. 
Conv1, 2, 3, 4 and 5 serve
to evaluate the energy of custom hardware, since they have a variety of input image sizes, channels, kernels, and convolution window sizes, and have results generated by others that we can use as reference points.


\subsection{General processor memory usage evaluation}

We evaluated both Westmere and Haswell processors.  Since the results are similar we will use the Xeon E5645 (Westmere) CPU as our base platform
for evaluating 
memory statistics on a general processor.
The system has 32KB~L1 data cache, 256K~L2 cache, 12MB~L3 cache, and runs at 2.4GHz. 
The memory usage is extracted by embedding PAPI, an x86 performance counter~\cite{browne2000portable}, 
into the CNN code.
To sanity check these numbers we 
compared the
PAPI results with 
the application running on
Zsim~\cite{sanchez2013zsim}, a cycle-accurate x86 architecture simulator, and the results were well correlated, within 10\% of each other.

Our blocked CNN algorithm uses
Ha\-lide~\cite{ragan2013halide, ragan2012decoupling}, a do\-main-specific language
for image processing, where blocking, vectorization, and multi-threading can be done using high level primitives.
%
%
We compare our blocked CNN implementation against two different versions of Caffe~\cite{Caffe}, one using  MKL linear algebra libraries~\cite{wang2014mkl} and one using ATLAS~\cite{whaley2001atlas}.

\subsection{Customized solution evaluation}
\label{sec:cusSol}
As mentioned in Section~\ref{sec:memEng}, we estimated access energy for memories of different sizes. The largest memories at the higher levels of the memory hierarchy are built from SRAM or DRAM depending on their size. For these memories we try to use wide bit widths and multi-banked memory to minimize energy cost. SRAMs become inefficient at small sizes, so the smallest buffers are created using a standard-cell register file generator. We also take advantage of the shifting window nature of many of the computations by building register files that can shift their data as well. This ability to shift data can be useful in stencil computations (\textit{i.e.,} convolution) when iterating in the $x$ dimension, since it allows us to load only the new column of data that is needed to update the image window rather than reloading the entire image patch.

We use
the register file generator in
the Cadence Xtensa Processor Generator~\cite{tensilica} to extract register file energy and area numbers. 
First we synthesize the design using Synopsys Design Compiler and a TSMC 45nm GS SVT library, and then place-and-route with the Synopsys IC Com\-pil\-er~\cite{synopsys}. We estimate the power of the post-routed designs using IC Compiler with activity factors extracted from Synopsys VCS simulations.

The energy and area of the computation datapath was generated in a similar way.
We built an arithmetic unit including 256 16-bit truncated multipliers, 16 reduction trees of adders, and 16 piece-wise linear approximation units for activation functions, similar to the core structure in DianNao~\cite{chen2014diannao}.
The arithmetic unit is pipelined and has a throughput of 256 MAC/cycle at 500MHz, reducing 16 inputs and 256 kernel values to 16 partial outputs.
We implement the design in RTL, and then synthesize and place-and-route using Synopsys Design Compiler and IC Compiler.

To evaluate SRAM area and energy, we derive parameters from CACTI~\cite{cacti_micro07} with multiple configurations targeting optimum energy/byte for various SRAM block sizes. The raw CACTI data was calibrated with results from a commercial 45nm memory compiler. The DRAM energy was 
estimated using information in a Micron tech note~\cite{micron_power}. 

We conservatively estimate the energy of communicating data, or doing scatter-gather operations, to be the access energy of a memory of the same size as the communicating hardware. The details are discussed in Section~\ref{sec:memEng}. 


\section{Experimental Results}

\label{sec:Results}


To validate the effectiveness of our blocking strategy, 
we first evaluate cache access behavior on a general processor. 
While our optimization aims to minimize memory energy, 
it also
minimizes 
cache accesses if the cache sizes are fixed.


Next, to emphasize its potential energy impact, we look at what effect our scheme could have in the highly energy-efficient custom architecture space.  We first look at energy considerations on single-core custom solutions, and then extend to multi-core.

\begin{figure}
   \centering
   \vspace{-5pt}
   \includegraphics[width=0.5\textwidth]{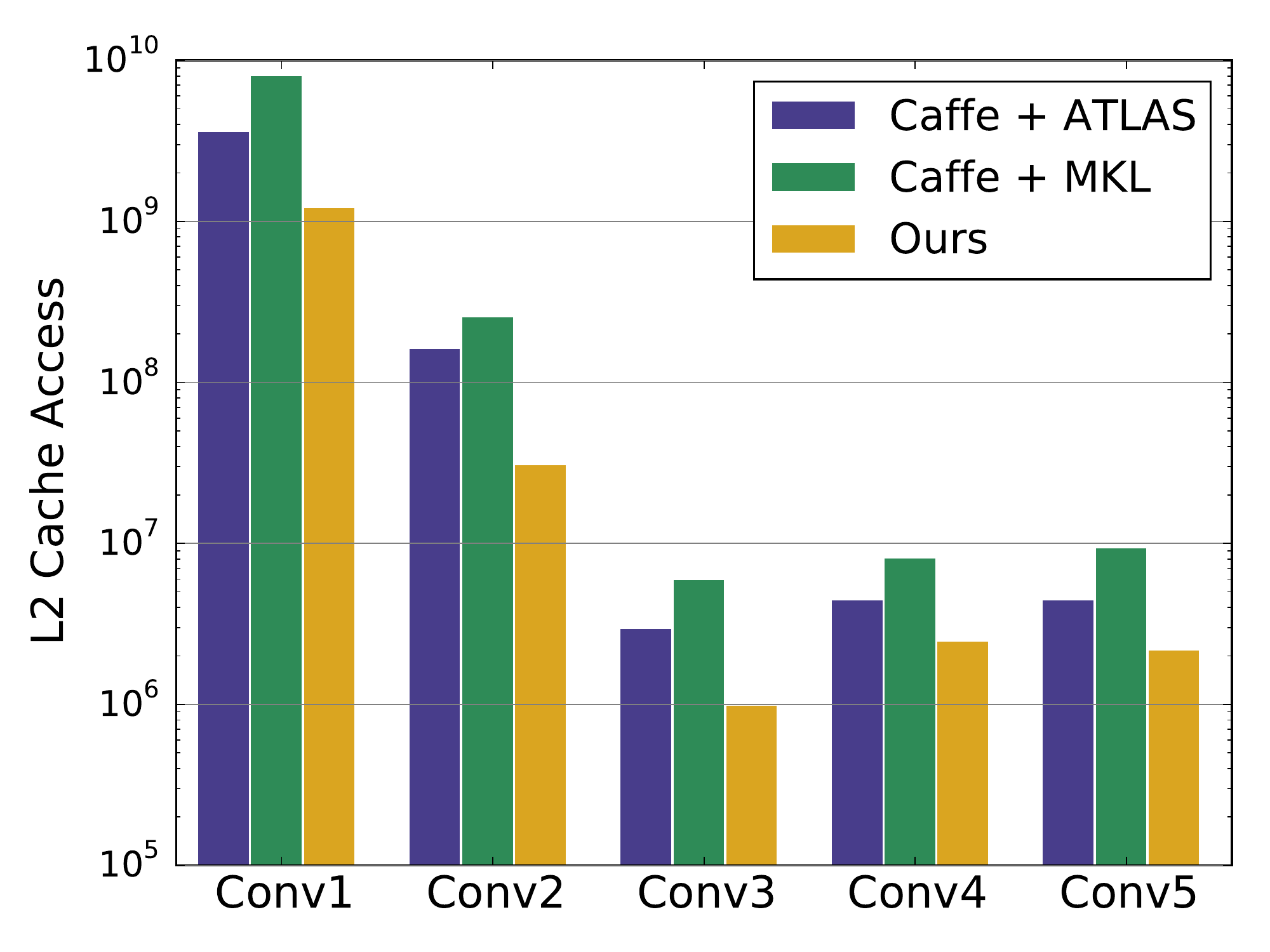} 
   \vspace{-15pt}
   \caption{L2 cache access counts (log scale) for two traditional BLAS-based implementations of Conv layers, versus our proposed blocking.}
   \label{fig:General_L1}
\end{figure}

\begin{figure}
   \centering
   \vspace{-20pt}
   \includegraphics[width=0.5\textwidth]{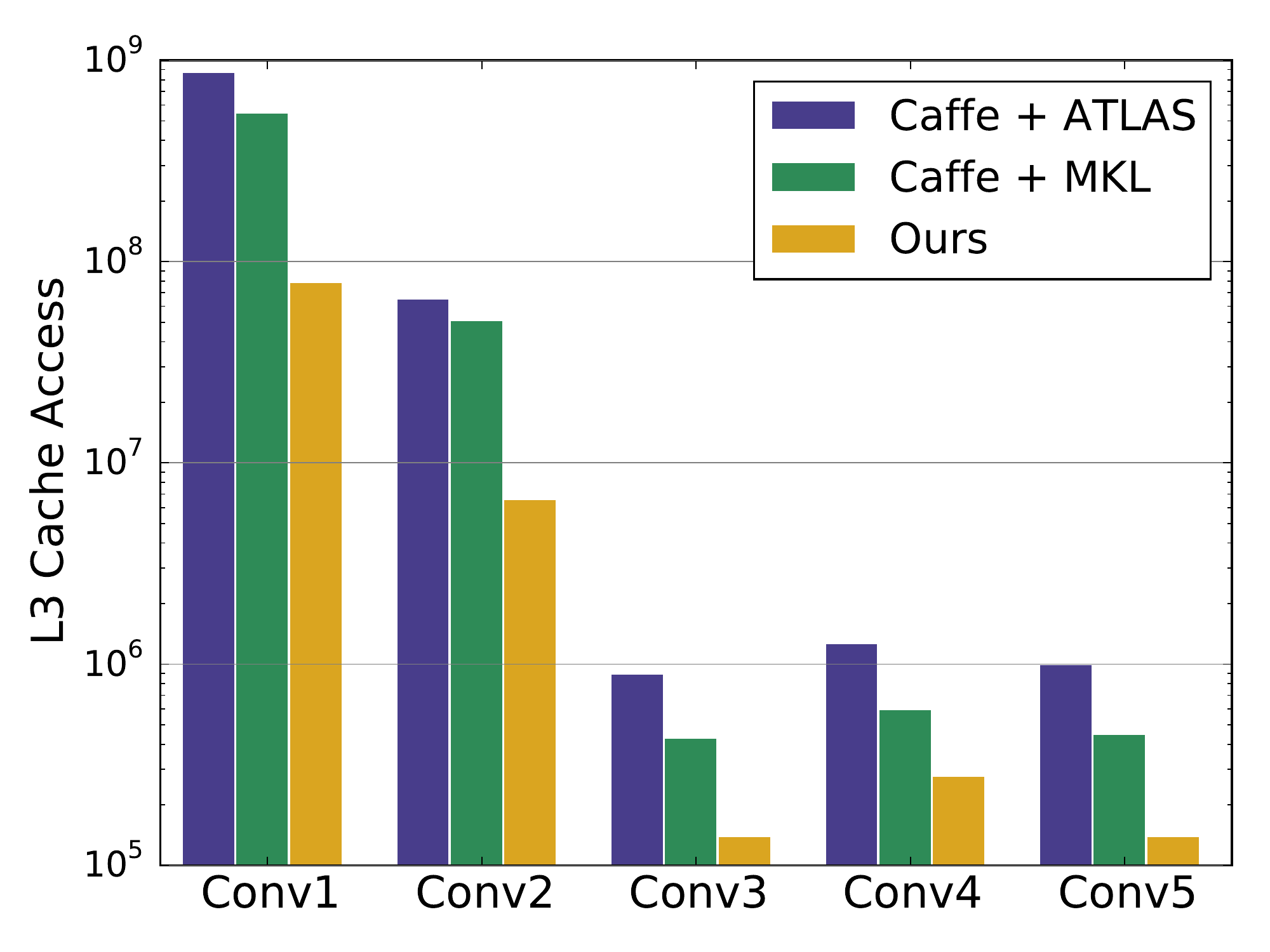} 
   \vspace{-15pt}
   \caption{L3 cache access counts (log scale) for two traditional BLAS-based implementations of Conv layers, versus our proposed blocking.}
   \label{fig:General_L2}
\end{figure}

\begin{figure*}
   \centering
   \includegraphics[width=1\textwidth]{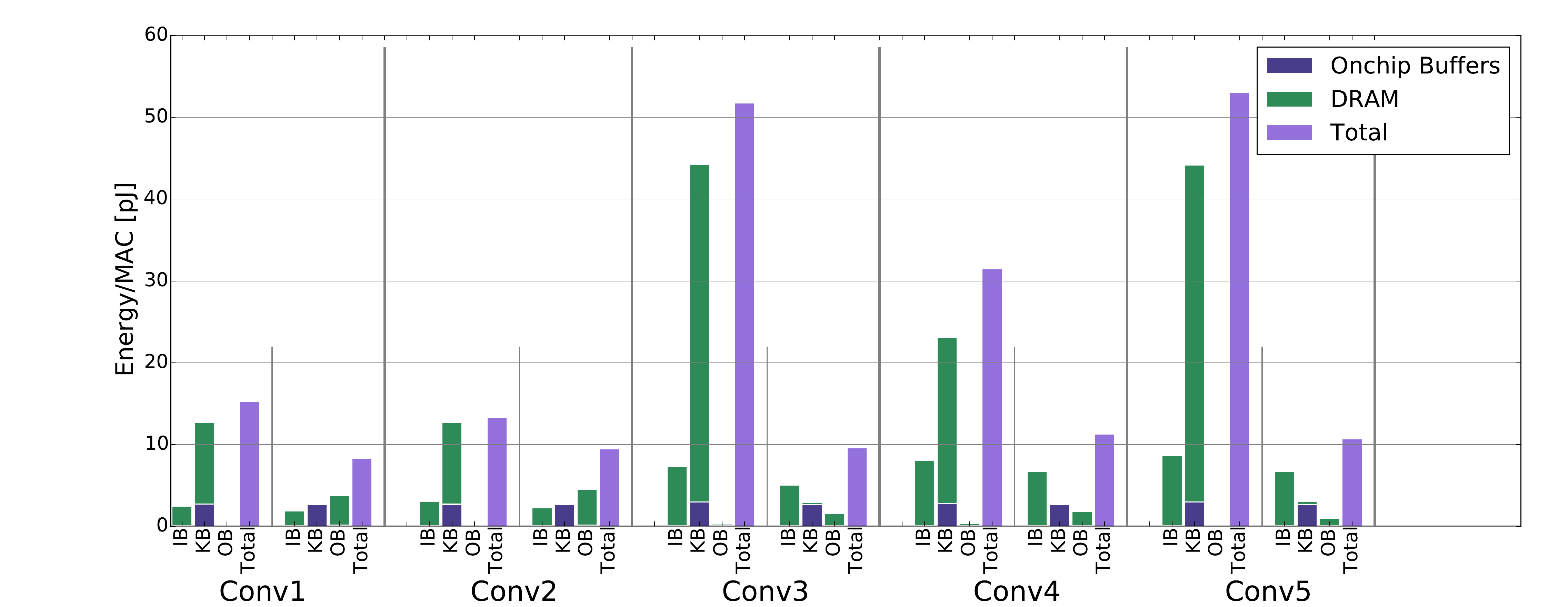} 
   \caption{IB, KB, OB and total energy, respectively,
   for benchmarks running on 
   DianNao,
   using the baseline schedule~(left) and the optimal schedule~(right). DRAM energy dominates the total memory energy, and the memory energy dwarfs the computation energy 
   (about 1 pJ/MAC).
   }
   \label{fig:Sched_Comp}
\end{figure*}

\subsection{General processor memory statistics}

To get a sense of how well our blocking optimization
is working, 
we first look at the cache statistics of the five Conv layers in our benchmarks.
Figures~\ref{fig:General_L1} and~\ref{fig:General_L2} respectively show the number of L2 and L3 cache accesses for various CNN implementations.    
In Figure~\ref{fig:General_L1}, 
our blocking
achieves the fewest L2 cache accesses on each of the five layers: 
accesses for the implementation using ATLAS are always greater than 2x and can be as high as 5x of our cache access. And the L2 cache access counts of the implementation using MKL are greater than 4x of ours, to a maximum of 8x.

In Figure~\ref{fig:General_L2}, 
our blocking
significantly reduces the L3 cache accesses for all benchmarks as well. For the five Conv layers, L3 cache accesses for the implementation using ATLAS are greater than 5x worse and can be as high as 11x of our cache access, with MKL ranging from 2x-7x worse.

%
%

Note that, for both L2 and L3, our blocking's
advantage decreases as it goes from Conv1 to Conv5.  The reason
is that the convolution window width and height
gradually decreases from 1 to 5.
This indicates that the later layers (Conv5) have a structure which better fits GEMM implementation than the earlier layers (Conv1).

Overall, the results illustrate that our blocking optimization is extremely effective in improving memory locality for fixed size memory. In addition, directly blocking the convolution layer achieves better memory usage than converting it to GEMM, and
this difference gets larger as the computation structures become more distinctive from GEMM. 

\newcolumntype{C}{>{\small\centering\arraybackslash}X}

\subsection{Custom core energy result} 
\label{sec:cusRes}

\begin{figure}
   \centering
   \includegraphics[
     width=0.48\textwidth,
   ]{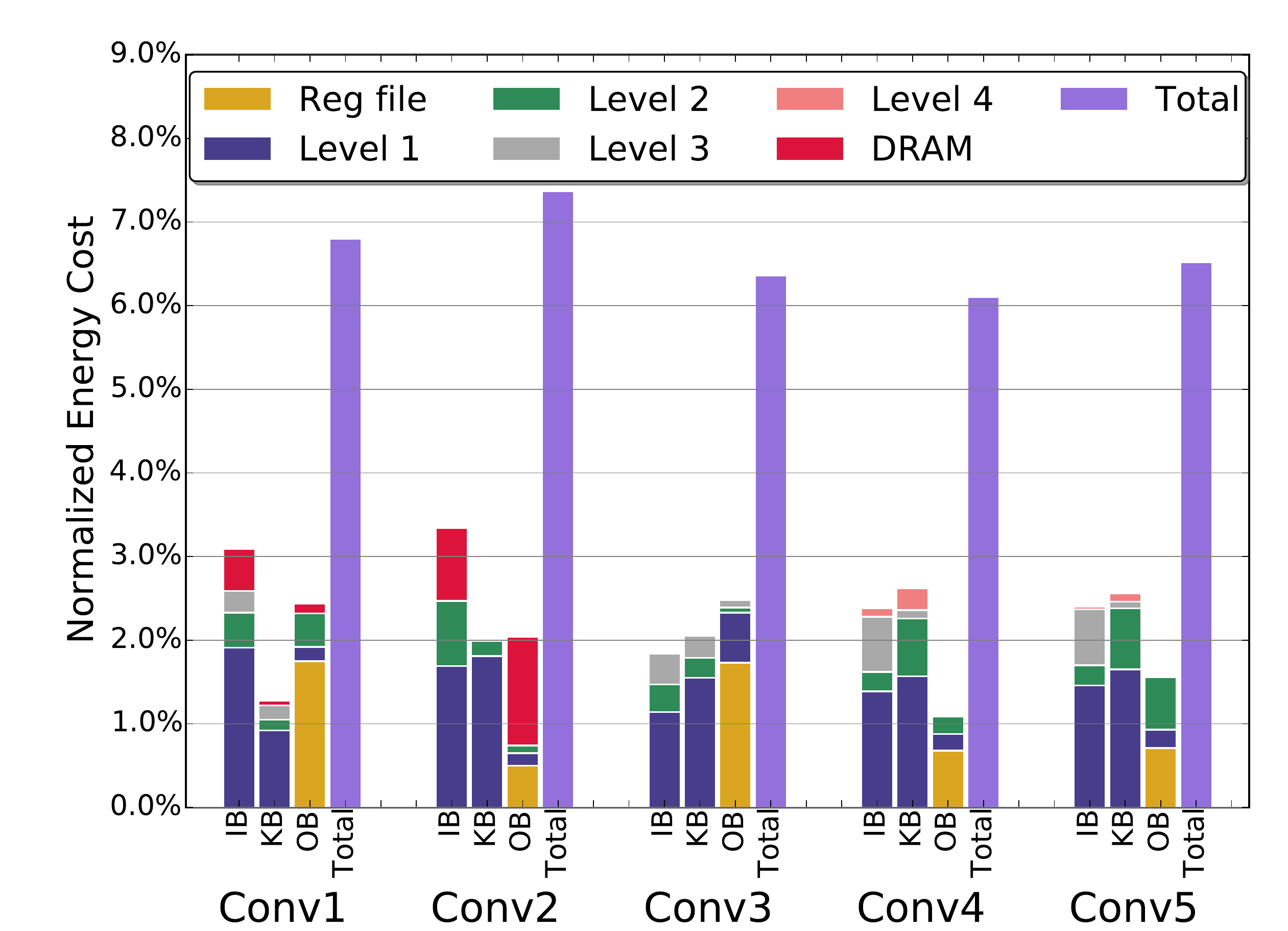} \tabularnewline
   \caption{IB, KB, OB and total energy, respectively, on the optimal architecture (core+memory) for five benchmarks, normalized by the total energy of each benchmark on DianNao's architecture with the optimal scheduling (Figure~\ref{fig:Sched_Comp}).
   }
   \label{fig:Mem_Comp_Norm}
\end{figure}

\begin{figure}
   \centering
   \includegraphics[
     width=0.48\textwidth,
   ]{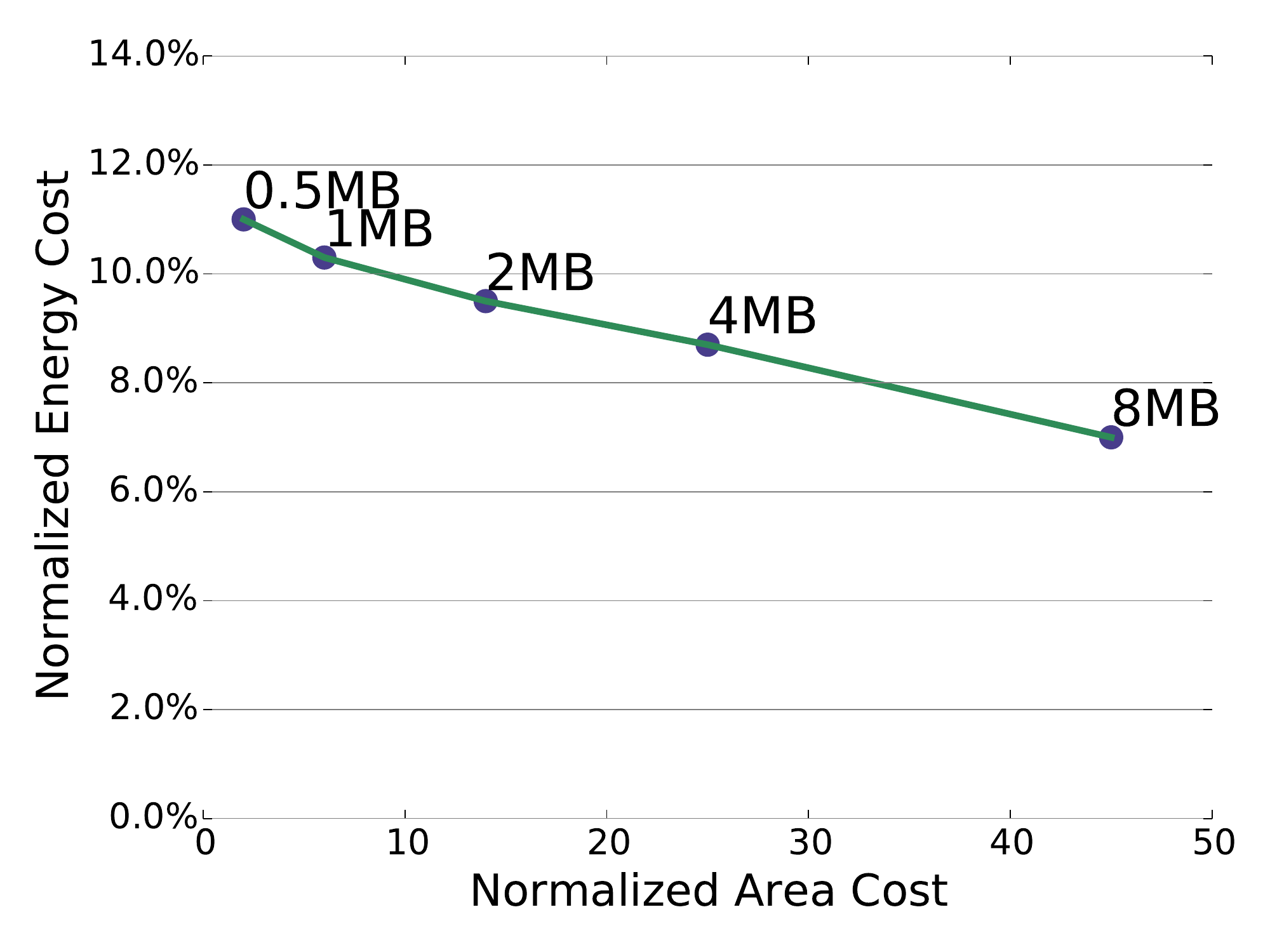} \tabularnewline
   \caption{Total energy and area for the optimal architecture, under various SRAM size constraints, normalized by the total energy and area of DianNao's baseline architecture, and using our optimal scheduling. The architecture with 1MB on-chip mem\-ory achieves about 10x energy reduction at the cost of 6x area.
   }
   \label{fig:Mem_Area_Limit}
\end{figure}

\begin{figure}[!ht]
   \centering
   \includegraphics[width=0.5\textwidth]{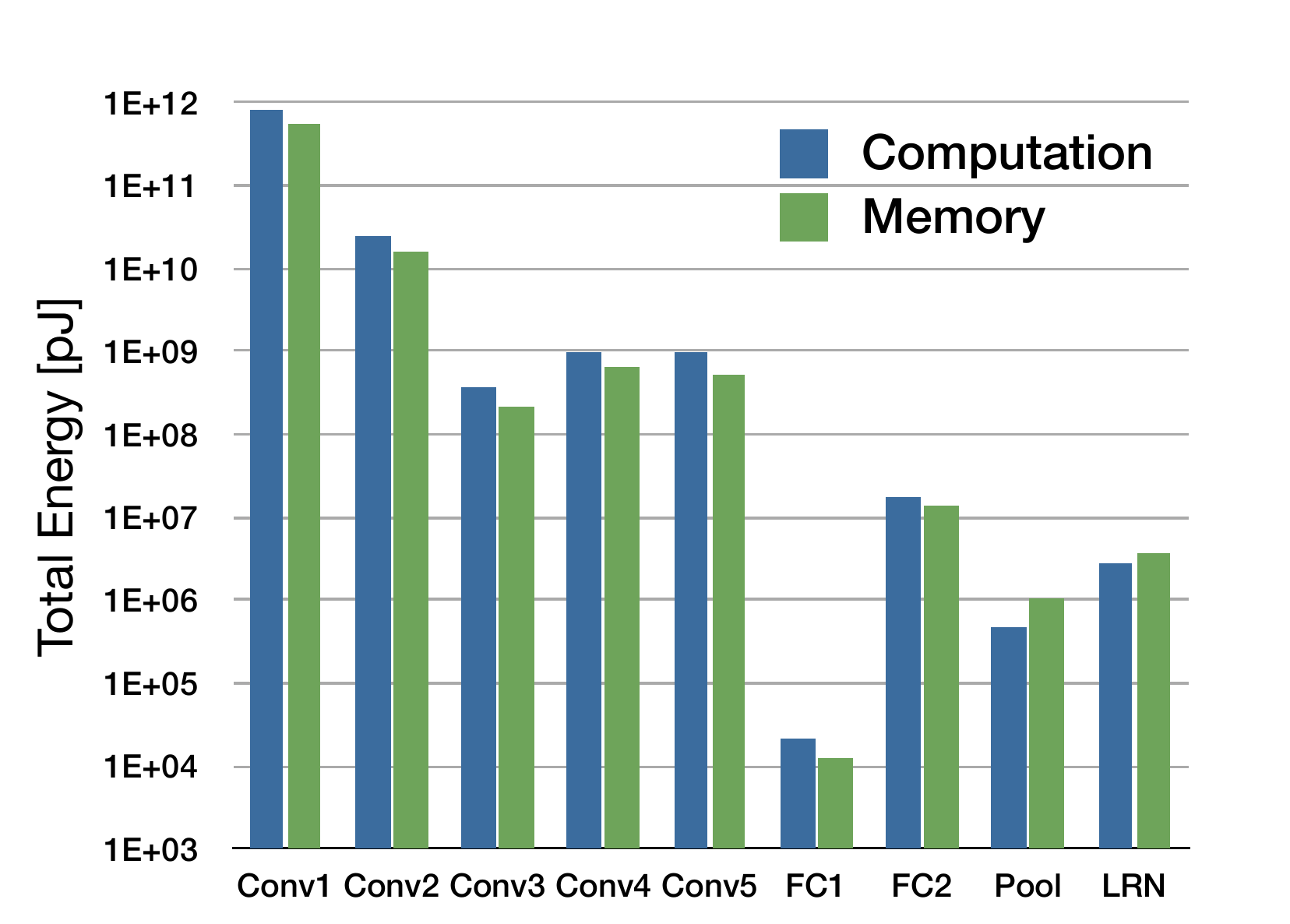} \tabularnewline
   \caption{Energy breakdown (log scale) of computation and memory for all benchmarks.
   Memory energy is less than 80\% of MAC energy for all convolutional and fully-connected layers. 
   }
   \label{fig:comp_memory}
   \vspace{8pt}
\end{figure}%

\begin{figure*}[!ht] 
   \centering
   \includegraphics
   [width=1\textwidth]
   {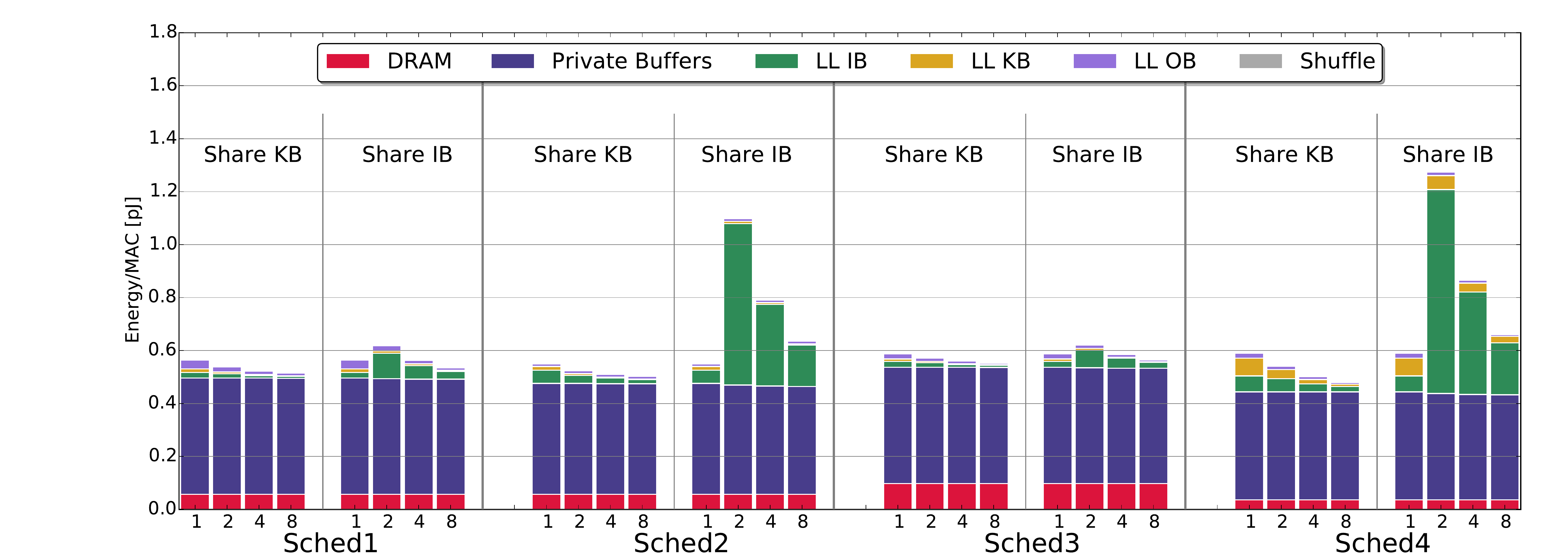} 
   \caption{Multi-core scaling of on-chip memory energy using shared KB and IB schemes 
   for Conv1 and its various sub-optimal schedules (sched1-4). 1, 2, 4, 8 represent the numbers of cores.
   \emph{Private buffers} represents the total memory energy spent inside a core, while \emph{LL IB}, \emph{LL KB} and \emph{LL OB} are the last-level buffers for the multiple core system. DRAM is the main memory energy cost. \emph{Shuffle} represents energy spent restoring the memory layout after computation.
   }
   \vspace{5pt}
   \label{fig:multi_core}
\end{figure*}

Having shown the benefits of blocking CNNs on a conventional CPU memory hierarchy, we next look at custom accelerators. To create a comparison point, we explore the effects of blocking on energy consumption in the DianNao
CNN accelerator~\cite{chen2014diannao}. 
DianNao
has separate on-chip SRAMs for IB, KB and OB (2KB, 32KB, 2KB respectively) and a 256-MAC core datapath.
For the baseline blocking scheme, we first used
DianNao's own pseudo-code,
but even the smallest IB ($K_0 \times C_{n}$) 
cannot fit in 2KB,
so all accesses to inputs go to DRAM.
So we
ended up blocking in the $x$ dimension once more  
to shrink the IB
to 2KB, reducing DRAM accesses by 4x.

Figure~\ref{fig:Sched_Comp} compares
this improved baseline schedule to the optimal scheduling found by our framework.
With the baseline schedule, the KB in DRAM consumes the most energy, especially for Conv3, 4 and~5, whose kernel sizes are large relative to the image size.
The optimized schedule improves kernel reuse by interchanging the loop order 
and reduces KB energy by 2x to 15x.
In addition, IB and OB DRAM energy dominates in the optimal schedule, and thus further energy reduction could be achieved by allocating more SRAM to IB and OB or even re-designing the memory hierarchy. 

To further reduce the energy we add this additional memory and co-design it
with the blocking schedule.
We assume the SRAM size is limited to 8MB to fit in a normal silicon die, while keeping the same core datapath design.
Figure~\ref{fig:Mem_Comp_Norm} demonstrates the energy reduction of optimal memory systems 
normalized to DianNao's one-level-buffer architecture. 
Even for the largest image sizes (Conv1 and Conv2), previous expensive IB and OB DRAM accesses
can be saved by exploiting the extra on-chip SRAM buffers. 
IB and OB in Conv2 still consume large DRAM energy
because the input and output images are large compared to kernels. 
The optimal design using an 8MB memory hierarchy improves energy efficiency by at least 13x, but it also
requires 45mm$^2$ which increases the required area 
by 45x vs. baseline DianNao.

If area is an issue, one can generate 
a curve showing
how the minimum energy changes versus area as in Figure~\ref{fig:Mem_Area_Limit}. 
For example, with a total memory of 1MB the energy efficiency now improves by 10x, and only requires 6x more chip area than DianNao. 
This large energy savings comes from
a combination of a large reduction in DRAM references, and being able to satisfy most references from small register-files.  Accomplishing both requires deep memory hierarchies.  
%

Figure~\ref{fig:comp_memory} depicts the energy consumption of memory accesses versus computation 
for different layers on our optimal 8MB system.
Our blocking scheme, along with a co-designed memory hierarchy,
drops the memory-access energy to computation energy ratio, originally 20x in the DianNao system, to less than 1x in our optimal system.
 
These results point out one limitation in using GPUs for these applications. Since they were
designed to keep their floating point functional units busy, they use large register-files to hold multiple runnable threads, and very high bandwidth memory systems with modest on-chip caching.  This increases the cost of operand fetch, both when the data is local in the register-file, and when it needs to be fetched from memory.  To gain efficiency, 
especially for the low precision needed for CNNs, 
GPUs will need to create deeper memory hierarchies, with smaller register-files near the compute units. 

\subsection{Multi-custom cores}

The large memory required for optimal blocking cannot be used with a single
core, since the resulting compute density would be too small. Instead, systems
with large memory will use multiple computing units. As discussed in 
Section~\ref{sec:Parallel}, there are  
two different schemes for parallelizing a problem on a multiple-core chip.
Figure~\ref{fig:multi_core} demonstrates how the two parallelization schemes, with different schedules and memory hierarchies, affect the energy efficiency of a system with up to eight cores. We chose the top four blocking schedules from the single core problem discussed in the previous section 
and we evaluated
the two parallelization methods for each of the four schedules as applied to layer Conv1.  

In all four schedules, the last level KB dominates the chip area.
If this large KB buffer is split between cores, then its energy will be reduced, but it means that the smaller IB must now be broadcast to all of the units.  Since this signal must travel a distance related to the total size of this memory, its energy now becomes as large as the large KB was. This means we 
still have a very large access energy (now the IB) for the partitioned KB buffer. The takeaway is that one should  
parallelize the hardware such that the large buffer is shared since it has the side effect of making the needed broadcast essentially free. With this unrolling, increasing the core count always gets better average energy since the other SRAM structures are partitioned and their energy/access reduces. 
Also since the ``shuffle" energy 
required to restore memory layout after computation 
seems to be small, it doesn't affect this decision. The net result is that when the right loop is unrolled in hardware, the performance can be increased with a small decrease in the energy per op.
\section{Conclusion}
\label{sec:Conclusion}

Modern computing systems are power limited, 
meaning
that performance can only improve by reducing
energy per operation. For convolutional neural networks, we show the large potential energy savings available from directly blocking the computation and provide a method for finding efficient schedules. When the memory system and loop schedule are co-optimized, the resulting machine can have energy/op dominated by the functional operations and not the memory system. To generate these optimized schedules, we build an analytical model and use it to either exhaustively or heuristically search possible schedules. This optimization procedure can be modified to find blocking schedules that optimize memory locality on a fixed memory hierarchy. Using this optimization mode we were able to improve the memory usage of software CNN implementations. 

We are now working to improve this framework to perform better register blocking for general processors, and to extend the framework for other computer vision applications. Once this is done we hope to integrate this work into Halide, to help programmers better block their code.
\vfill





\bibliographystyle{IEEEtranS}
\bibliography{ref}

\begin{thebibliography}{10}
\providecommand{\url}[1]{#1}
\csname url@samestyle\endcsname
\providecommand{\newblock}{\relax}
\providecommand{\bibinfo}[2]{#2}
\providecommand{\BIBentrySTDinterwordspacing}{\spaceskip=0pt\relax}
\providecommand{\BIBentryALTinterwordstretchfactor}{4}
\providecommand{\BIBentryALTinterwordspacing}{\spaceskip=\fontdimen2\font plus
\BIBentryALTinterwordstretchfactor\fontdimen3\font minus
  \fontdimen4\font\relax}
\providecommand{\BIBforeignlanguage}[2]{{%
\expandafter\ifx\csname l@#1\endcsname\relax
\typeout{** WARNING: IEEEtranS.bst: No hyphenation pattern has been}%
\typeout{** loaded for the language `#1'. Using the pattern for}%
\typeout{** the default language instead.}%
\else
\language=\csname l@#1\endcsname
\fi
#2}}
\providecommand{\BIBdecl}{\relax}
\BIBdecl

\bibitem{abuzaid2015caffe}
F.~Abuzaid, S.~Hadjis, C.~Zhang, and C.~R{\'e}, ``{C}affe con {T}roll: Shallow
  ideas to speed up deep learning,'' \emph{arXiv preprint arXiv:1504.04343},
  2015.

\bibitem{bergstra2010deep}
J.~Bergstra, F.~Bastien, J.~Turian, R.~Pascanu, O.~Delalleau, O.~Breuleux,
  P.~Lamblin, G.~Desjardins, D.~Erhan, Y.~Bengio \emph{et~al.}, ``Deep learning
  on {GPU}s with {T}heano,'' in \emph{The Learning Workshop-Research
  Abstract-Oral preferred (Feb. 18, 2010)}, 2010.

\bibitem{bondhugula2016pluto+}
U.~Bondhugula, A.~Acharya, and A.~Cohen, ``The pluto+ algorithm: A practical
  approach for parallelization and locality optimization of affine loop
  nests,'' \emph{ACM Trans. On Programming Languages and Systems (TOPLAS)},
  2016.

\bibitem{bondhugula2014tiling}
U.~Bondhugula, V.~Bandishti, A.~Cohen, G.~Potron, and N.~Vasilache, ``Tiling
  and optimizing time-iterated computations on periodic domains,'' in
  \emph{Proceedings of the 23rd international conference on Parallel
  architectures and compilation}.\hskip 1em plus 0.5em minus 0.4em\relax ACM,
  2014, pp. 39--50.

\bibitem{bondhugula2008practical}
U.~Bondhugula, A.~Hartono, J.~Ramanujam, and P.~Sadayappan, ``A practical
  automatic polyhedral parallelizer and locality optimizer,'' in
  \emph{Proceedings of the 29th ACM SIGPLAN Conference on Programming Language
  Design and Implementation}, ser. PLDI '08.\hskip 1em plus 0.5em minus
  0.4em\relax New York, NY, USA: ACM, 2008, pp. 101--113.

\bibitem{browne2000portable}
S.~Browne, J.~Dongarra, N.~Garner, G.~Ho, and P.~Mucci, ``A portable
  programming interface for performance evaluation on modern processors,''
  \emph{International Journal of High Performance Computing Applications},
  vol.~14, no.~3, pp. 189--204, 2000.

\bibitem{tensilica}
{Cadence Design Systems, Inc.}, ``Tensilica customizable processor {IP},''
  \url{http://ip.cadence.com/ipportfolio/tensilica-ip}.

\bibitem{chen2014diannao}
T.~Chen, Z.~Du, N.~Sun, J.~Wang, C.~Wu, Y.~Chen, and O.~Temam, ``{DianNao:} a
  small-footprint high-throughput accelerator for ubiquitous
  machine-learning,'' in \emph{Proc. 19th Int'l Conf. on Architectural Support
  for Programming Languages and Operating Systems}.\hskip 1em plus 0.5em minus
  0.4em\relax ACM, 2014, pp. 269--284.

\bibitem{chen2014dadiannao}
Y.~Chen, T.~Luo, S.~Liu, S.~Zhang, L.~He, J.~Wang, L.~Li, T.~Chen, Z.~Xu,
  N.~Sun \emph{et~al.}, ``{DaDianNao:} a machine-learning supercomputer,'' in
  \emph{47th Annual IEEE/ACM Int'l Symp. on Microarchitecture (MICRO)}.\hskip
  1em plus 0.5em minus 0.4em\relax IEEE, 2014, pp. 609--622.

\bibitem{dongarra1990set}
J.~J. Dongarra, J.~Du~Croz, S.~Hammarling, and I.~S. Duff, ``A set of level 3
  basic linear algebra subprograms,'' \emph{ACM Transactions on Mathematical
  Software (TOMS)}, vol.~16, no.~1, pp. 1--17, 1990.

\bibitem{dundar12teradeep}
A.~Dundar, J.~Jin, V.~Gokhale, B.~Martini, and E.~Culurciello, ``Memory access
  optimized routing scheme for deep networks on a mobile coprocessor,''
  \emph{Algorithms}, vol.~12, p.~15, 2014.

\bibitem{farabet2011neuflow}
C.~Farabet, B.~Martini, B.~Corda, P.~Akselrod, E.~Culurciello, and Y.~LeCun,
  ``{NeuFlow:} a runtime reconfigurable dataflow processor for vision,'' in
  \emph{IEEE Computer Society Conf. on Computer Vision and Pattern Recognition
  Workshops (CVPRW)}.\hskip 1em plus 0.5em minus 0.4em\relax IEEE, 2011, pp.
  109--116.

\bibitem{frigo2005cache}
M.~Frigo and V.~Strumpen, ``Cache oblivious stencil computations,'' in
  \emph{Proceedings of the 19th annual international conference on
  Supercomputing}.\hskip 1em plus 0.5em minus 0.4em\relax ACM, 2005, pp.
  361--366.

\bibitem{gokhale2014teradeep}
V.~Gokhale, J.~Jin, A.~Dundar, B.~Martini, and E.~Culurciello, ``A 240
  {G}-ops/s mobile coprocessor for deep neural networks,'' in \emph{IEEE Conf.
  on Computer Vision and Pattern Recognition Workshops (CVPRW)}.\hskip 1em plus
  0.5em minus 0.4em\relax IEEE, 2014, pp. 696--701.

\bibitem{Goto:2008:AHM:1356052.1356053}
K.~Goto and R.~A. {v}an~{d}e {G}eijn, ``Anatomy of high-performance matrix
  multiplication,'' \emph{ACM Trans. Math. Softw.}, vol.~34, no.~3, pp.
  12:1--12:25, May 2008.

\bibitem{Gunnels:2001:FHM:645455.653765}
J.~A. Gunnels, G.~M. Henry, and R.~A. {v}an~{d}e {G}eijn, ``A family of
  high-performance matrix multiplication algorithms,'' in \emph{Proceedings of
  the International Conference on Computational Sciences-Part I}, ser. ICCS
  '01.\hskip 1em plus 0.5em minus 0.4em\relax London, UK, UK: Springer-Verlag,
  2001, pp. 51--60.

\bibitem{hinton2006fast}
G.~Hinton, S.~Osindero, and Y.-W. Teh, ``A fast learning algorithm for deep
  belief nets,'' \emph{Neural computation}, vol.~18, no.~7, pp. 1527--1554,
  2006.

\bibitem{micron_power}
M.~T. Inc., ``{TN-41-01: Calculating Memory System Power for DDR3},''
  \url{http://www.micron.com/support/power-calc}, 2007.

\bibitem{jarrett2009best}
K.~Jarrett, K.~Kavukcuoglu, M.~Ranzato, and Y.~LeCun, ``What is the best
  multi-stage architecture for object recognition?'' in \emph{12th Int'l Conf.
  on Computer Vision}.\hskip 1em plus 0.5em minus 0.4em\relax IEEE, 2009, pp.
  2146--2153.

\bibitem{Caffe}
Y.~Jia, E.~Shelhamer, J.~Donahue, S.~Karayev, J.~Long, R.~Girshick,
  S.~Guadarrama, and T.~Darrell, ``Caffe: Convolutional architecture for fast
  feature embedding,'' in \emph{Proceedings of the ACM International Conference
  on Multimedia}.\hskip 1em plus 0.5em minus 0.4em\relax ACM, 2014, pp.
  675--678.

\bibitem{jin2014teradeep}
J.~Jin, V.~Gokhale, A.~Dundar, B.~Krishnamurthy, B.~Martini, and
  E.~Culurciello, ``An efficient implementation of deep convolutional neural
  networks on a mobile coprocessor,'' in \emph{IEEE 57th Int'l Midwest Symp. on
  Circuits and Systems (MWSCAS)}.\hskip 1em plus 0.5em minus 0.4em\relax IEEE,
  2014, pp. 133--136.

\bibitem{krizhevsky2010convolutional}
A.~Krizhevsky and G.~Hinton, ``Convolutional deep belief networks on
  {CIFAR-10},'' \emph{Unpublished manuscript}, 2010.

\bibitem{NIPS2012_4824}
A.~Krizhevsky, I.~Sutskever, and G.~E. Hinton, ``{ImageNet} classification with
  deep convolutional neural networks,'' in \emph{Advances in Neural Information
  Processing Systems 25}, F.~Pereira, C.~Burges, L.~Bottou, and K.~Weinberger,
  Eds.\hskip 1em plus 0.5em minus 0.4em\relax Curran Associates, Inc., 2012,
  pp. 1097--1105.

\bibitem{DBLP:journals/corr/Lavin15}
A.~Lavin, ``{maxDNN}: An efficient convolution kernel for deep learning with
  maxwell {GPU}s,'' \emph{CoRR}, vol. abs/1501.06633, 2015.

\bibitem{liu2015pudiannao}
D.~Liu, T.~Chen, S.~Liu, J.~Zhou, S.~Zhou, O.~Teman, X.~Feng, X.~Zhou, and
  Y.~Chen, ``{PuDianNao:} a polyvalent machine learning accelerator,'' in
  \emph{Proc. 20th Int'l Conf. on Architectural Support for Programming
  Languages and Operating Systems}.\hskip 1em plus 0.5em minus 0.4em\relax ACM,
  2015, pp. 369--381.

\bibitem{cacti_micro07}
N.~Muralimanohar, R.~Balasubramonian, and N.~Jouppi, ``{Optimizing NUCA
  Organizations and Wiring Alternatives for Large Caches with CACTI 6.0},'' in
  \emph{Proceedings of the 40th Annual IEEE/ACM International Symposium on
  Microarchitecture}, ser. MICRO 40.\hskip 1em plus 0.5em minus 0.4em\relax
  Washington, DC, USA: IEEE Computer Society, 2007, pp. 3--14.

\bibitem{cuDNN}
NVIDIA, ``{cuDNN},'' \url{https://developer.nvidia.com/cuDNN}, 2014.

\bibitem{peemen2013memory}
M.~Peemen, A.~A. Setio, B.~Mesman, and H.~Corporaal, ``Memory-centric
  accelerator design for convolutional neural networks,'' in \emph{31st Int'l
  Conf. on Computer Design (ICCD)}.\hskip 1em plus 0.5em minus 0.4em\relax
  IEEE, 2013, pp. 13--19.

\bibitem{pham2012neuflow}
P.-H. Pham, D.~Jelaca, C.~Farabet, B.~Martini, Y.~LeCun, and E.~Culurciello,
  ``{NeuFlow:} dataflow vision processing system-on-a-chip,'' in \emph{55th
  Int'l Midwest Symp. on Circuits and Systems (MWSCAS)}.\hskip 1em plus 0.5em
  minus 0.4em\relax IEEE, 2012, pp. 1044--1047.

\bibitem{pouchet2013polyhedral}
L.-N. Pouchet, P.~Zhang, P.~Sadayappan, and J.~Cong, ``Polyhedral-based data
  reuse optimization for configurable computing,'' in \emph{Proceedings of the
  ACM/SIGDA international symposium on Field programmable gate arrays}.\hskip
  1em plus 0.5em minus 0.4em\relax ACM, 2013, pp. 29--38.

\bibitem{ragan2012decoupling}
J.~Ragan-Kelley, A.~Adams, S.~Paris, M.~Levoy, S.~P. Amarasinghe, and
  F.~Durand, ``Decoupling algorithms from schedules for easy optimization of
  image processing pipelines.'' \emph{ACM Trans. Graph.}, vol.~31, no.~4,
  p.~32, 2012.

\bibitem{ragan2013halide}
J.~Ragan-Kelley, C.~Barnes, A.~Adams, S.~Paris, F.~Durand, and S.~Amarasinghe,
  ``Halide: a language and compiler for optimizing parallelism, locality, and
  recomputation in image processing pipelines,'' \emph{ACM SIGPLAN Notices},
  vol.~48, no.~6, pp. 519--530, 2013.

\bibitem{sanchez2013zsim}
D.~Sanchez and C.~Kozyrakis, ``Zsim: Fast and accurate microarchitectural
  simulation of thousand-core systems,'' in \emph{Proceedings of the 40th
  Annual International Symposium on Computer Architecture}, ser. ISCA
  '13.\hskip 1em plus 0.5em minus 0.4em\relax New York, NY, USA: ACM, 2013, pp.
  475--486.

\bibitem{sermanet2011traffic}
P.~Sermanet and Y.~LeCun, ``Traffic sign recognition with multi-scale
  convolutional networks,'' in \emph{Neural Networks (IJCNN), The 2011
  International Joint Conference on}.\hskip 1em plus 0.5em minus 0.4em\relax
  IEEE, 2011, pp. 2809--2813.

\bibitem{simonyan2014vggnet}
K.~Simonyan and A.~Zisserman, ``Very deep convolutional networks for
  large-scale image recognition,'' \emph{arXiv preprint arXiv:1409.1556}, 2014.

\bibitem{strigl5452452}
D.~Strigl, K.~Kofler, and S.~Podlipnig, ``Performance and scalability of
  {GPU}-based convolutional neural networks,'' in \emph{Parallel, Distributed
  and Network-Based Processing (PDP), 2010 18th Euromicro International
  Conference on}, Feb 2010, pp. 317--324.

\bibitem{strzodka2010cache}
R.~Strzodka, M.~Shaheen, D.~Pajak, and H.-P. Seidel, ``Cache oblivious
  parallelograms in iterative stencil computations,'' in \emph{Proceedings of
  the 24th ACM International Conference on Supercomputing}.\hskip 1em plus
  0.5em minus 0.4em\relax ACM, 2010, pp. 49--59.

\bibitem{synopsys}
{Synopsys, Inc.}, ``Synopsys,'' \url{http://www.synopsys.com}.

\bibitem{turaga2010convolutional}
S.~C. Turaga, J.~F. Murray, V.~Jain, F.~Roth, M.~Helmstaedter, K.~Briggman,
  W.~Denk, and H.~S. Seung, ``Convolutional networks can learn to generate
  affinity graphs for image segmentation,'' \emph{Neural Computation}, vol.~22,
  no.~2, pp. 511--538, 2010.

\bibitem{wang2014mkl}
E.~Wang, Q.~Zhang, B.~Shen, G.~Zhang, X.~Lu, Q.~Wu, and Y.~Wang, ``Intel math
  kernel library,'' in \emph{High-Performance Computing on the Intel Xeon
  Phi}.\hskip 1em plus 0.5em minus 0.4em\relax Springer, 2014, pp. 167--188.

\bibitem{whaley2001atlas}
R.~C. Whaley, A.~Petitet, and J.~J. Dongarra, ``Automated empirical
  optimizations of software and the atlas project,'' \emph{Parallel Computing},
  vol.~27, no.~1, pp. 3--35, 2001.

\bibitem{zhang2015optimizing}
C.~Zhang, P.~Li, G.~Sun, Y.~Guan, B.~Xiao, and J.~Cong, ``Optimizing
  {FPGA}-based accelerator design for deep convolutional neural networks,'' in
  \emph{Proceedings of the 2015 ACM/SIGDA International Symposium on
  Field-Programmable Gate Arrays}.\hskip 1em plus 0.5em minus 0.4em\relax ACM,
  2015, pp. 161--170.

\end{thebibliography}


\end{document}